\documentclass[12pt,letterpaper]{article}
\input{epsf}
\usepackage[pdftex]{graphicx,color}
\usepackage{hyperref}
\input{psfig}
\usepackage{amsmath,amssymb}
\usepackage{fancyhdr}
\newcommand\ltap{\
  \raise.3ex\hbox{$<$\kern-.75em\lower1ex\hbox{$\sim$}}\ }
\newcommand\gtap{\
  \raise.3ex\hbox{$>$\kern-.75em\lower1ex\hbox{$\sim$}}\ }

\newcommand\simge{\mathrel{%
   \rlap{\raise 0.511ex \hbox{$>$}}{\lower 0.511ex \hbox{$\sim$}}}}
\newcommand\simle{\mathrel{
   \rlap{\raise 0.511ex \hbox{$<$}}{\lower 0.511ex \hbox{$\sim$}}}}

\newcommand{\slashchar}[1]%
        {\kern .25em\raise.18ex\hbox{$/$}\kern-.70em #1}
\def\lsim{\mathrel{\raise.3ex\hbox{$<$\kern-.75em\lower1ex\hbox{$\sim$}}}}
\def\gsim{\mathrel{\raise.3ex\hbox{$>$\kern-.75em\lower1ex\hbox{$\sim$}}}}

\newcommand\CL{{\cal L}}
\newcommand\CM{{\cal M}}

\newcommand\CO{{\cal O}}

\newcommand\ul{\underline}
\newcommand\ol{\overline}
\newcommand\be{\begin{equation}}
\newcommand\ee{\end{equation}}
\newcommand\bea{\begin{eqnarray}}
\newcommand\eea{\end{eqnarray}}
\newcommand\ba{\begin{array}}
\newcommand\ea{\end{array}}
\newcommand\nn{\nonumber}

\newcommand{\half}{\ensuremath{\frac{1}{2}}}

\newcommand{\thalf}{\textstyle{\frac{1}{2}}}

\newcommand{\tfourth}{\textstyle{\frac{1}{4}}}

\newcommand{\teighth}{\textstyle{\frac{1}{8}}}

\newcommand\dagg{\dagger}

\newcommand\gev{{\rm GeV}}
\newcommand\tev{{\rm TeV}}

\newcommand\pb{{\rm pb}}

\newcommand\fb{{\rm fb}}
\newcommand\ifb{{\rm fb}^{-1}}

\newcommand\ellm{\ell^-}

\newcommand\ellp{\ell^+}

\newcommand\cbeta{c_\beta}
\newcommand\sbeta{s_\beta}

\newcommand{\LGW}{\Lambda_{\rm GW}}

\newcommand{\bM}{\ol{M}}

\newcommand{\Hpt}{H^{\prime\,2}}

\begin{document}

\title{
\vspace{-50mm}
{\Large{\bf Higgs alignment and the top quark}}\\
\medskip
} \author{ {\large Estia J.~Eichten$^{1}$\thanks{eichten@fnal.gov}}\, and
  Kenneth Lane$^{2}$\thanks{lane@bu.edu}\\
{\large {$^{1}$}Theoretical Physics Group, Fermi National Accelerator
  Laboratory}\\
{\large P.O. Box 500, Batavia, Illinois 60510}\\
{\large $^{2}$Department of Physics, Boston University}\\
{\large 590 Commonwealth Avenue, Boston, Massachusetts 02215}\\
} \maketitle

\vspace{-1.0cm}

\begin{abstract}
 
  There is a surprising connection between the top quark and Higgs alignment
  in Gildener-Weinberg multi-Higgs doublet models. Were it not for the top
  quark and its large mass, the coupling of the $125\,\gev$ Higgs boson $H$
  to gauge bosons and fermions would be indistinguishable from those of the
  Standard Model Higgs. The top quark's coupling to a single Higgs doublet
  breaks this perfect alignment in higher orders of the Coleman-Weinberg loop
  expansion of the effective potential. But the effect is still small,
  $\simle \CO(1\%)$, and probably experimentally inaccessible.

  \end{abstract}


\newpage

\section*{I. Introduction}
\begin{figure}[ht!]
 \begin{center}
\includegraphics[width=2.65in, height=2.65in]{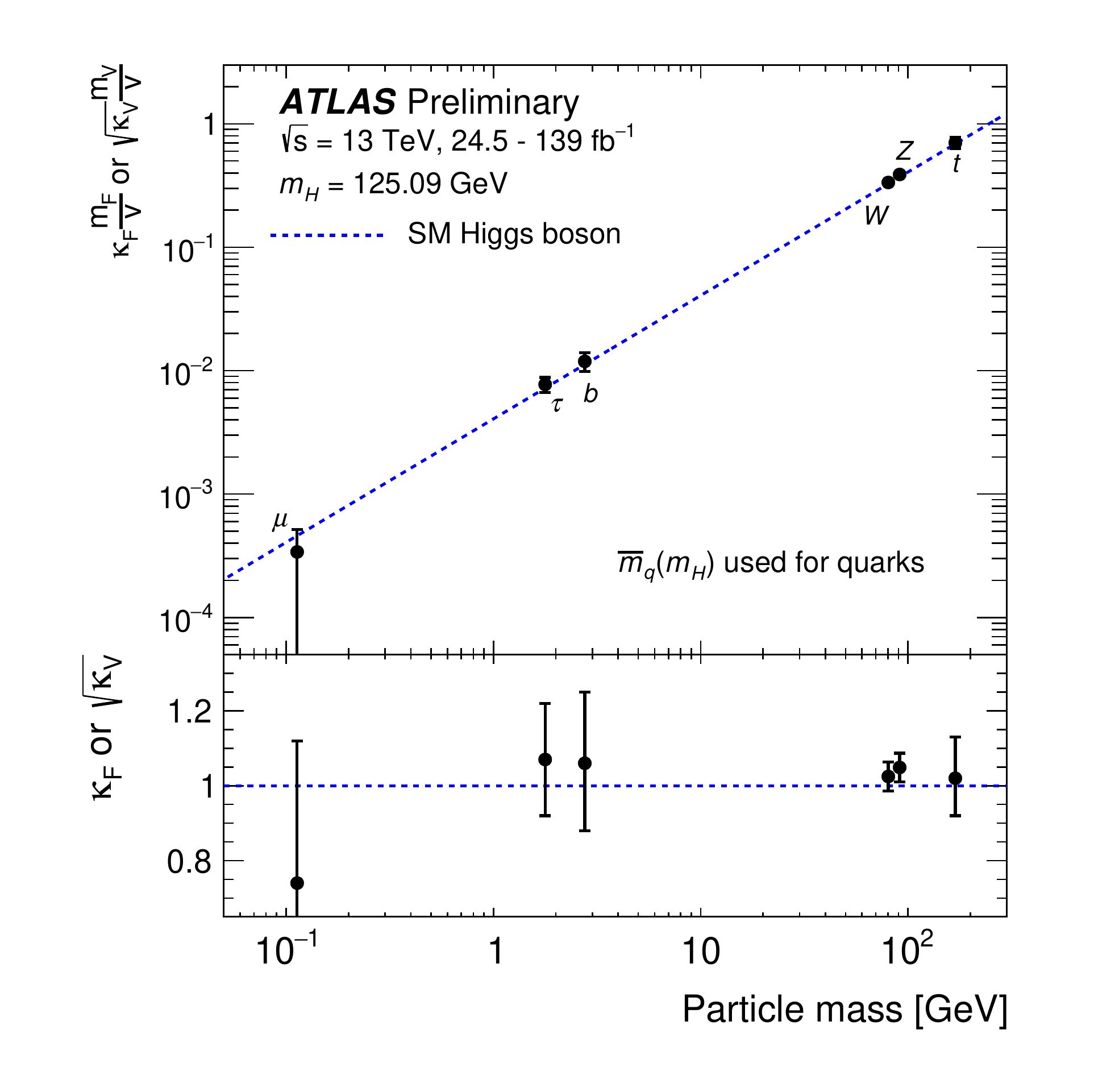}
\includegraphics[width=2.65in, height=2.65in]{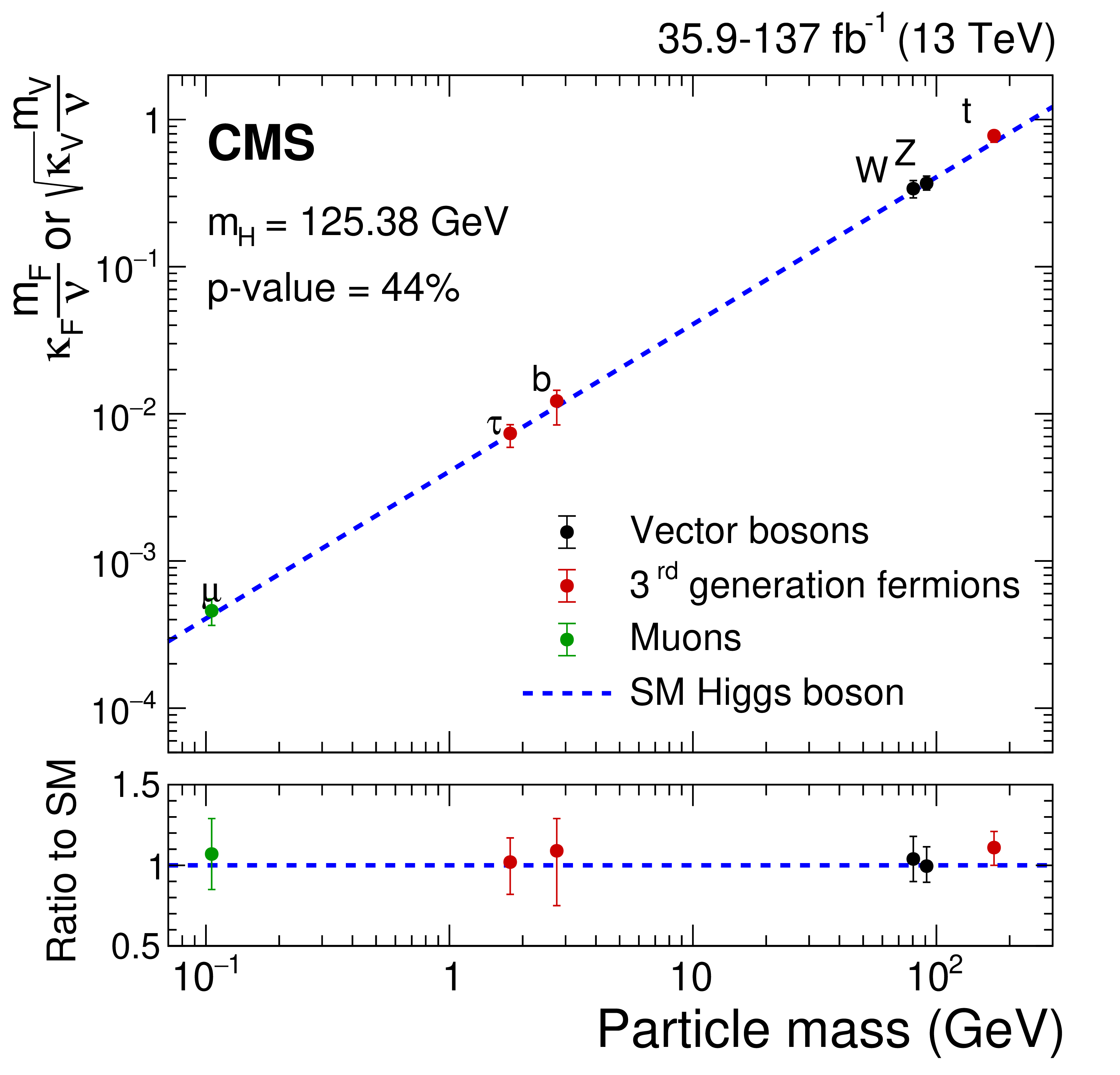}
\caption{The mass-dependent couplings of quarks, leptons and the $W$ and $Z$
  as measured by
  ATLAS~({\url{https://atlas.web.cern.ch/Atlas/GROUPS/PHYSICS/CombinedSummaryPlots/HIGGS/}}
  and CMS
  {\url{https://cms-results.web.cern.ch/cms-results/public-results/publications/HIG/index.html}}.}
  \label{fig:SMHiggs}
 \end{center}
 \end{figure}
 The $125\,\gev$ Higgs boson $H$ discovered at the LHC in
 2012~\cite{Aad:2012tfa,Chatrchyan:2012ufa} is consistent in all measurements
 with the single Higgs boson of the Standard Model~(SM). This is dramatically
 illustrated in Fig.~\ref{fig:SMHiggs} where the couplings of fermions and
 weak gauge bosons to $H$ as measured by ATLAS and CMS are plotted. This
 degree of agreement is puzzling. Many well-motivated attempts to
 cure the problems of the SM --- most famously, naturalness --- require two
 or more Higgs multiplets. Why, then, does $H$ have SM couplings? The usual
 answer is ``Higgs
 alignment''~\cite{Boudjema:2001ii,Gunion:2002zf,Carena:2013ooa,
   Haber:2018ltt}. However, with a few exceptions that rely on elaborate
 global symmetries or supersymmetry~\cite{Ivanov:2007de,Dev:2014yca,
   Benakli:2018vqz,Darvishi:2019ltl}, implementations of alignment suffer
 large radiative corrections.

 In Gildener-Weinberg (GW) multi-Higgs models of electroweak (EW) symmetry
 breaking~\cite{Gildener:1976ih}, the classical Lagrangian is scale-invariant
 --- so that the Higgs potential is purely quartic and fermions acquire mass
 only from Higgs boson vacuum expectation values. At tree level, $H$ is a
 Goldstone boson of spontaneously broken scale symmetry. And, in this
 approximation, $H$ {\em naturally} has the same structure as the Goldstone
 bosons eaten by $W^\pm$ and $Z^0$. In $N$-Higgs-doublet models (NHDMs),
\be\label{eq:Phii}
\Phi_i = \frac{1}{\sqrt{2}}\left(\ba{c}\sqrt{2} \phi_i^+ \\ \rho_i + i
  a_i \ea\right), \quad i = 1,2,\dots N,
\ee
these Goldstone bosons are
\be\label{EWGB}
w^\pm = \sum_{i=1}^{N} v_i\phi_i^\pm/v, \quad z = \sum_{i=1}^{N} v_i a_i/v,
\ee
where $v_i$ is the VEV of the $C\!P$-even scalar~$\rho_i$ and
$v = \sqrt{\sum_{i=1}^{N} v_i^2}$. The Higgs boson is
\be\label{Halign}
H = \sum_{i=1}^{N} v_i\rho_i/v.
\ee
Thus, $H$ has exactly the same couplings to EW gauge bosons and to fermions
(and, hence, to the gluon and the photon) as the single Higgs boson of the
Standard Model; i.e., $H$ is aligned.

In Sec.~II of this paper, we show that, but for the top quark, this alignment
would be perfect through second order in the Coleman-Weinberg loop expansion
of the effective potential~\cite{Coleman:1973jx}. The top quark's presence
upsets perfect alignment, but only by a small amount, at most $\CO(1\%)$. In
Sec.~III we discuss the experimental consequences of this alignment. In
short, experimental searches for new Higgs bosons, such as $H',A$ and
$H^\pm$, via weak vector boson fusion or decay and Drell-Yan production in
association with $H$, will remain fruitless. We also update the promising
paths to discovery of these new Higgses at the LHC. They rely on the fact
that these new bosons must lie below 400--500~GeV.

\section*{II. Higgs alignment in the GW-2HDM}

We discuss the top quark's role in Higgs alignment in the context of an
$N = 2$~Higgs doublet model introduced by Lee and Pilaftsis in
2012~\cite{Lee:2012jn}. However, by the Glashow-Weinberg criterion that all
quarks of a given electric charge must couple to a single Higgs doublet to
avoid flavor-changing neutral current interactions mediated by neutral Higgs
exchange~\cite{Glashow:1976nt}, our conclusion is true in any GW-NHDM. This
GW-2HDM was updated in 2018~\cite{Lane:2018ycs} to make it consistent with
LHC data at the time. The modification used the following $\mathbb{Z}_2$
symmetry on the Higgs doublets and fermions:
\be\label{eq:Z2}
\Phi_1 \to -\Phi_1,\,\, \Phi_2 \to \Phi_2, \quad
\psi_L \to -\psi_L,\,\, \psi_{uR} \to \psi_{uR},\,\,
\psi_{dR} \to \psi_{dR}.
\ee
This is the usual type-I 2HDM~\cite{Branco:2011iw}, but with $\Phi_1$ and
$\Phi_2$ interchanged. (The effect of this is that experimental lower limits
on $\tan\beta = v_2/v_1$ in other type-I models are lower limits on
$\cot\beta$ in this model.)  The most important experimental constraint came
from CMS~\cite{Khachatryan:2015qxa} and ATLAS~\cite{Aaboud:2018cwk} searches
for charged Higgs decay into $t\bar b$.  Consistency with those experiments
required $\tan\beta \simle 0.5$ for $M_{H^\pm} \simle 500\,\gev$. This is
discussed in Sec.~III.

The GW tree-level potential for this model is purely
quartic~\cite{Gildener:1976ih} so that, since all masses in the model arise
from Higgs VEVs, the Lagrangian is scale-invariant at this level:
\bea\label{eq:Vzero}
V_0(\Phi_1,\Phi_2) &=&\lambda_1 (\Phi_1^\dagg \Phi_1)^2 +
\lambda_2 (\Phi_2^\dagg \Phi_2)^2 +
\lambda_3(\Phi_1^\dagg \Phi_1)(\Phi_2^\dagg \Phi_2)\nn \\
&+& \lambda_4(\Phi_1^\dagg \Phi_2)(\Phi_2^\dagg \Phi_1)+
\thalf\lambda_5\left((\Phi_1^\dagg \Phi_2)^2 + (\Phi_2^\dagg
  \Phi_1)^2\right).
\eea
The five quartic couplings~$\lambda_i$ in Eq.~(\ref{eq:Vzero}) are real and,
so, $V_0$ is $C\!P$-invariant. The couplings $\lambda_{1,2} > 0$ for
positivity of the potential.

The trivial minimum of $V_0$ occurs at $\Phi_1 = \Phi_2 = 0$. But a nontrivial
flat minimum of $V_0$ can occur on the ray
\be\label{eq:theray}
\Phi_{1\beta} = \frac{1}{\sqrt{2}} \left(\ba{c} 0\\ \phi\,\cbeta
  \ea\right),\quad
\Phi_{2\beta} = \frac{1}{\sqrt{2}} \left(\ba{c} 0\\ \phi\,\sbeta \ea\right),
\ee
where $\cbeta = \cos\beta$, $\sbeta = \sin\beta$ with $\beta \neq 0,\pi/2$ a
fixed angle and $0 < \phi < \infty$ a real mass scale.\footnote{It is
  easily proved that any such purely quartic potential as well as its first
  derivatives vanish at {\em{\ul{any}}} extremum so that
  $V_0(\Phi_{i\beta}) = 0$~\cite{Lane:2019dbc}.} The nontrivial extremal
conditions are
\be\label{eq:lamconds}
\lambda_1 + \thalf\lambda_{345} \tan^2\beta =
\lambda_2 + \thalf\lambda_{345} \cot^2\beta = 0,
\ee
where $\lambda_{345} = \lambda_3 + \lambda_4 + \lambda_5 < 0$ for positivity
of $V_0$. Eqs.~(\ref{eq:lamconds}) hold in all orders of the loop expansion
of the effective potential~\cite{Gildener:1976ih}; this is important in our
subsequent development (see
Eqs.~(\ref{eq:extrx},\ref{eq:extrx1},\ref{eq:extrx2})). This extremum
spontaneously (but not explicitly) breaks scale invariance, as well as the EW
gauge symmetry, and $H$ is the corresponding Goldstone boson.

Following Ref.~\cite{Lane:2019dbc}, we use the ``aligned basis'' of the Higgs
fields because the scalars' mass matrices will remain very nearly diagonal in
that basis beyond the tree approximation (also see
Ref.~\cite{Haber:2018ltt}). That is the essence of Higgs alignment in GW
models and, in this and similar models, it is broken, but only slightly, by
the top quark. This basis is:
\bea\label{eq:aligned}
\Phi &=& \Phi_1\cbeta + \Phi_2\sbeta = \frac{1}{\sqrt{2}} \left(\ba{c}
  \sqrt{2}w^+\\ H + iz\ea\right), \nn\\\\
\Phi' &=& -\Phi_1\sbeta  + \Phi_2\cbeta = \frac{1}{\sqrt{2}} \left(\ba{c}
  \sqrt{2}H^+\\ H' + iA \ea\right).\nn
\eea
On the ray Eq.~(\ref{eq:theray}) on which $V_0$ has nontrivial extrema, these
fields are $\Phi = (0,\phi)/\sqrt{2}$ and $\Phi'=0$. In this basis,
$H,z,w^\pm$ are massless and unmixed with the $H',A,H^\pm$ whose ``masses''
are
\be\label{eq:mevec}
M^2_{H'} = -\lambda_{345}\phi^2,\,\,\,M^2_A =
-\lambda_5\phi^2,\,\,\,M^2_{H^\pm} = -\thalf\lambda_{45}\phi^2.
\ee
Thus, the flat potential is indeed a minimum on the ray~(\ref{eq:theray})
(albeit degenerate with the trivial one) if, like $\lambda_{345}$,
$\lambda_5$ and $\lambda_{45} =\lambda_4 + \lambda_5$ are negative.

To establish the top quark's role in Higgs alignment of GW-NHDMs, it suffices
to consider this model in one-loop order of the effective potential,
$V_0 + V_1$. This potential provides a lower minimum than $V_0 = 0$ by
picking out a particular value $v$ of $\phi$, explicitly breaking the
scale symmetry of $V_0$, and giving $H$ a nonzero mass. The one-loop
potential is~\cite{Martin:2001vx}
\be\label{eq:Vone} V_1 = \frac{1}{64\pi^2}\sum_n \alpha_n\bM_n^4
\left(\ln\frac{\bM_n^2}{\LGW^2} - k_n\right).
\ee
Only very massive particles contribute to $V_1$. They are
$n = (W^\pm,Z,t,H',A,H^\pm)$ in this model. The constants are
$\alpha_n = (6,3,-12,1,1,2)$; $k_n = 5/6$ for the weak gauge bosons and~3/2
for scalars and the top-quark.\footnote{$V_1$ was calculated in the Landau
  gauge using the ${\ol{\rm MS}}$ renormalization scheme.}
The background-field-dependent masses $\bM_n^2$
in Eq.~(\ref{eq:Vone}) are~\cite{Jackiw:1974cv,Lee:2012jn}
\be\label{eq:bkgM}
\bM_n^2 = \left\{\ba{l} M_n^2\left(2 \left(\Phi^\dagg \Phi +
      \Phi^{\prime\,\dagg} \Phi'\right)/\phi^2\right) = 
   M_n^2 \left(H^2 + H^{\prime\,2} + \cdots\right)/\phi^2, \quad n \neq t\\
    M_t^2 \left(2\Phi_1^\dagg\Phi_1/(\phi\cbeta)^2\right) =
    M_t^2\left((H\cbeta - H'\sbeta)^2 + \cdots\right)/(\phi\cbeta)^2,\\
    \hspace{3.75cm} = M_t^2\left((H - H'\tan\beta)^2 + \cdots\right)/\phi^2
    \ea\right.
\ee
where $M_n^2
\propto \phi^2$ is the actual squared ``mass'' of
particle~$n$.
The form of $\bM_t^2$
is dictated by the type-I coupling of fermions to the $\Phi_1$
doublet in Eq.(\ref{eq:Z2}). This difference controls the breaking of
Higgs alignment through second order in the loop expansion. The
renormalization scale $\LGW$ will be fixed relative to the Higgs VEV $v
= 246\,\gev$ in Eq.~(\ref{eq:lgw1}) below.

The one-loop extremal conditions are~\cite{Gildener:1976ih}
\be\label{eq:extrx}
\left.\frac{\partial(V_0+V_1)}{\partial H}
  \right\vert_{\langle\,\rangle + \delta_1 H + \delta_1 H'} =
\left.\frac{\partial (V_0+V_1)}{\partial H'}
  \right\vert_{\langle\,\rangle  + \delta_1 H + \delta_1 H'} = 0.
\ee
%
Here, we follow GW's analysis by expanding around the tree-level VEVs
$\langle H\rangle = \phi$, $\langle H'\rangle = 0$ while allowing for
$\CO(V_1)$ shifts $\delta_1H$ and $\delta_1H'$ 
in those VEVs --- and from perfect Higgs alignment. 
Recall that the tree-level extremal conditions $(\partial
V_0/\partial H)_{\langle\,\rangle} = (\partial V_0/\partial
H')_{\langle\,\rangle} = 0$ remain in force.
To $\CO(V_1)$ this expansion results in
\bea
\label{eq:extrx1}
&& \left.\frac{\partial V_1}{\partial H}\right\vert_{\langle\,\rangle} =
\frac{1}{16\pi^2 v} \sum_{n}\alpha_n
M_n^4\left(\ln\frac{M_n^2}{\LGW^2} +\half - k_n\right) = 0,\hspace{1.0cm}\\
\label{eq:extrx2}
&& \left.\frac{\partial^2
    V_0}{\partial\Hpt}\right\vert_{\langle\,\rangle}\delta_1H' +
\left.\frac{\partial V_1}{\partial H'}\right\vert_{\langle\,\rangle} =
M^2_{H'}\,\delta_1 H' -\frac{\alpha_t M_t^4\tan\beta}{16\pi^2 v}
\left(\ln\frac{M_t^2}{\LGW^2} +\half - k_t\right) = 0,\hspace{1.0cm} \eea
where, by Eq.~(\ref{eq:bkgM}), the first derivative with respect to $H'$
of the $n\neq t$ terms in $V_1$ vanish because they are quadratic in $H'$.
Eq.~(\ref{eq:extrx1}) provides a definition of the renormalization scale
$\LGW$ in terms of the VEV $\phi = v$ at which the minimum of
$V_1$ occurs. It can be rewritten as~\cite{Gildener:1976ih}
\be
\label{eq:lgw1}
0 = \sum_{n}\alpha_n M_n^4\left(\ln\frac{M_n^2}{\LGW^2} +\half - k_n\right)
  = A + {\thalf} B + B\ln\left(\frac{v^2}{\LGW^2}\right),
\ee
where $A = \sum_n \alpha_n M_n^4 (\ln(M_n^2/v^2) - k_n)$ and
$B = \sum_n \alpha_n M_n^4$, so that
$\ln(\LGW^2/v^2) = A/B +{\thalf}$.\footnote{As discussed in
  Ref.~\cite{Gildener:1976ih}, Eq.~(\ref{eq:extrx1}) does not lead to a
  minimum of $V_1$ unless $B > 0$. With the known masses of $W^\pm,Z,t$ and
  $H$, $B>0$; see Eq.~(\ref{eq:M0sq1}).} Note that $M_n^2 \propto v^2$ so
that $\LGW/v$ is a function of coupling constants only.

From Eq.~(\ref{eq:extrx2}), the shift $\delta_1H'$ in $\langle H'\rangle$ is
given by the tadpole formula:
\be\label{eq:del1Hpr}
\delta_1H'=  -\frac{1}{M^2_{H'}}\left.\frac{\partial V_1}{\partial
    H'}\right\vert_{\langle\,\rangle} =
 \frac{\alpha_t M_t^4\tan\beta}{16\pi^2 M^2_{H'}v} 
 \left(\ln\frac{M_t^2}{\LGW^2} +\half - k_t\right).
\ee
As an example of its magnitude, we take $M_{H'} = 400\,\gev$,
$\LGW = 260\,\gev$ and $\tan\beta = 0.5$. Then $\delta_1H' = 1.57\,\gev$
which, when added in quadrature with $v=246\,\gev$, amounts to an increase of
$0.002\,\%$.\footnote{Because $\delta_1H$ is not determined in $\CO(V_1)$, we
  can set it to zero here. This is consistent with our expectation that
  $\delta H = \CO(\delta^2)$ where $\delta = \CO(V_1)$ is the $H$--$H'$
  mixing angle.} 

Eq.~(\ref{eq:del1Hpr}) establishes the connection of the top quark to Higgs
alignment: The large mass of the top quark ensures its appearance in the
effective potential $V_1$ while the Glashow-Weinberg
criterion~\cite{Glashow:1976nt} implies
$(\partial \bM_t^2/\partial H')_{\langle\,\rangle} \neq 0$; hence the small
$\CO(V_1)$ shift away from perfect alignment. 
The elements of the $C\!P$-even mass matrix $\CM^2_{0^+}$ in $\CO(V_1)$
further emphasize this connection:
\bea
\label{eq:M0sq1}
\CM_{HH}^2 &=& \left.\frac{\partial^2 V_1}{\partial
    H^2}\right\vert_{\langle\,\rangle} = 
\frac{1}{8\pi^2 v^2}\sum_n \alpha_n M_n^4,\\
%
\label{eq:M0sq2}
\CM_{HH'}^2 &=& \left.\frac{\partial^3 V_0}{\partial H\partial
    \Hpt}\right\vert_{\langle\,\rangle}\delta_1H' +
    \left.\frac{\partial^2 V_1}{\partial H\partial
        H'}\right\vert_{\langle\,\rangle}\nn\\
&=& -\frac{\alpha_t M_t^4\tan\beta}{16\pi^2 v^2} \left(\ln
  \frac{M_t^2}{\LGW^2} + \frac{5}{2} - k_t\right), \\
\label{eq:M0sq3}
\CM_{H'H'}^2 &=& \left.\frac{\partial^2
    V_0}{\partial\Hpt}\right\vert_{\langle\,\rangle} + \left.\frac{\partial^3
    V_0}{\partial H^{\prime\,3}}\right\vert_{\langle\,\rangle}\delta_1H' +
     \left.\frac{\partial^2
         V_1}{\partial\Hpt}\right\vert_{\langle\,\rangle}\nn\\
%
%
&=& M^2_{H'} + \frac{\alpha_t M_t^4}{8\pi^2 v^2}\left(\ln\frac{M_t^2}{\LGW^2} +
  \frac{1}{2} - k_t + \tan^2\beta\right).
%
\eea
At this level, only the top quark prevents $\CM^2_{0^+}$ being diagonal and
the Higgs boson being completely aligned.

To repeat: Because the Glashow-Weinberg criterion applies to {\em any} EW
model in which quarks of a given charge acquire all their mass from the
scalars, the top quark's role in Higgs alignment holds in any GW-NHDM. The
additional complications of the two-loop effective potential do not alter
this conclusion.\footnote{E.~J.~Eichten and K.~Lane, in preparation.}

\section*{III. Experimental consequences}

ATLAS and CMS discovered the 125~GeV Higgs $H$ relatively easily because of
its rather strong coupling to weak bosons: production via $WW$ and $ZZ$
fusion and decay to $WW^*$ and $ZZ^*$. It's worth remembering that, despite
its lower statistics, $H \to ZZ^* \to 4$~leptons was much more convincing at
first than $H \to \gamma\gamma$. Furthermore, $gg$ fusion of $H$ via the
top-quark loop was important because the $\bar ttH$ coupling has its {\em
  full-strength} value, $M_t/v$. Because of this success, it seems, a
lamp-post strategy has been adopted for many searches of
Beyond-Standard-Model (BSM) Higgs bosons. This is especially true for heavier
neutral Higgses such as~$H'$ and~$A$. The web abounds with searches for $H'$
and $A$ production via weak-boson fusion or gluon fusion followed by their
decay to weak boson pairs or to $ZH$, as well as
$W^\pm Z \to H^\pm \to W^\pm Z$ or $W^\pm H$; see, e.g.,
{\url{https://twiki.cern.ch/twiki/bin/view/AtlasPublic}} and
{\url{https://cms-results.web.cern.ch/cms-results/public-results/publications/HIG/SUS.html}}.

All of these processes are strongly hindered by Higgs alignment in GW
models.\footnote{Similarly suppressed are the production in future lepton
  colliders of BSM scalars singly-produced in
  $e^+e^-,\,\mu^+\mu^- \to Z \to ZH'/A$ or via weak-boson fusion.}  This fact
is codified in the (non-Goldstone) Higgs bosons' interactions with
electroweak bosons and fermions. They are taken from Ref.~\cite{Lane:2018ycs}
but, because the $H$--$H'$ mixing angle
$\delta \simle \CO(1\%)$~\cite{Lane:2019dbc}, so that the rates for
Higgs-alignment-violating processes are suppressed by at least a factor
of~$10^{-4}$, it is more illuminating to write them in terms of the
aligned-basis fields $H$ and $H'$ rather than the mass eigenstates
$H_1 = H\cos\delta - H'\sin\delta$ and $H_2 = H\sin\delta + H'\cos\delta$.
\bea\label{eq:PhiEW}
\CL_{EW} &=&  ie H^- \overleftrightarrow{\partial_\mu} H^+
              \left(A^\mu + Z^\mu \cot 2\theta_W \right)
+ \frac{e}{\sin 2\theta_W} \left(H'\overleftrightarrow{\partial_\mu}A \right)
Z^\mu \nn\\
&+& \frac{ig}{2}\left(H^+
  \overleftrightarrow{\partial_\mu}(H'+iA) W^{-\,\mu} -
  H^- \overleftrightarrow{\partial_\mu}(H' - iA)W^{+\,\mu}\right) \nn\\
&+& H \left(gM_W\, W^{+\,\mu} W^{-}_\mu +
       {\thalf}\sqrt{g^2 + g^{\prime\,2}}M_Z\, Z^\mu Z_\mu\right) + \nn\\
&+& \left(H^2+H^{\prime\,2} +A^2\right)\left({\tfourth}g^2\, W^{+\,\mu}
W^{-}_\mu +{\teighth}(g^2+g^{\prime\,2})\,Z^\mu Z_\mu\right)\nn\\
&+& H^+H^- \left(e^2(A_\mu + Z_\mu\cot 2\theta_W)^2 +
{\tfourth}g^2 \,W_\mu^+ W^{-\,\mu}\right).
\eea
\begin{eqnarray}\label{eq:yukawa}
{\cal{L}}_Y &=& \frac{\sqrt{2}\tan\beta}{v}
      \sum_{k,l=1}^3\left[H^+\left(\bar u_{kL} V_{kl}\,m_{d_l}d_{lR}
      -\bar u_{kR}\, m_{u_k} V_{kl}\, d_{lL} +
      m_{\ell_k}\bar\nu_{kL}\ell_{kR}\,\delta_{kl}\right) + {\rm
      h.c.}\right] \nonumber\\
   &-& \left(\frac{v + H - H'\tan\beta}{v} \right)
       \sum_{k=1}^3 \left(m_{u_k} \bar u_k u_k + m_{d_k} \bar d_k d_k
                         +m_{\ell_k}\bar\ell_k \ell_k\right) \nonumber\\
   &-& \frac{i A\tan\beta}{v} \sum_{k=1}^3 \left(m_{u_k} \bar u_k \gamma_5 u_k
     - m_{d_k} \bar d_k\gamma_5 d_k - m_{\ell_k}\bar\ell_k
     \gamma_5\ell_k\right),
\end{eqnarray}
where $V = U_L^\dagg D_L$ is the CKM matrix. Note that gluon fusion and
two-photon decay of $H'$ and $A$ via a top-quark loop are suppressed by
$\tan^2\beta < 0.25$~\cite{Lane:2018ycs}. So long as $\tan^2\beta$ is not
much smaller, this can, in principle, be overcome by the data expected in
Run~3. And, so long as most BSM Higgs decays are to fermion pairs, the
$\tan^2\beta$ suppression does not affect decay branching ratios.

The prospects for testing the GW-2HDM (and similar models) are much brighter
than these comments suggest.\footnote{The following discussion is updated
  from Ref.~\cite{Lane:2019dbc} to include analyses using the full LHC Run~2
  data set of $139\,\ifb$.} They stem from the fact that the one-loop
approximation for the Higgs boson's mass in Eq.~(\ref{eq:M0sq1}) implies a
sum rule for the BSM Higgs masses~\cite{Lee:2012jn,Hashino:2015nxa,
  Lane:2018ycs,Braathen:2020vwo}:
\be\label{eq:gensum}
\left(M^4_{H'} + M^4_A + 2M^4_{H^\pm}\right)^{1/4} = 540\,\gev.
\ee
Eq.~(\ref{eq:gensum}) tells us that the BSM scalars are lighter than
400--500~GeV.\footnote{Because of limits from LHC searches for charged
  Higgses lighter than $M_t$ and neutral Higgses lighter than $M_H$ (see the
  ATLAS and CMS websites noted above), it is likely that they are heavier
  than $\sim 180\,\gev$.} In using this sum rule, we shall assume that
$M_A = M_{H^\pm}$, an assumption justified by the fact that it makes the BSM
scalars' contribution to the $T$-parameter vanish
identically~\cite{Battye:2011jj,Pilaftsis:2011ed}.)

\begin{figure}[h!]
\includegraphics[width=1.0\textwidth]{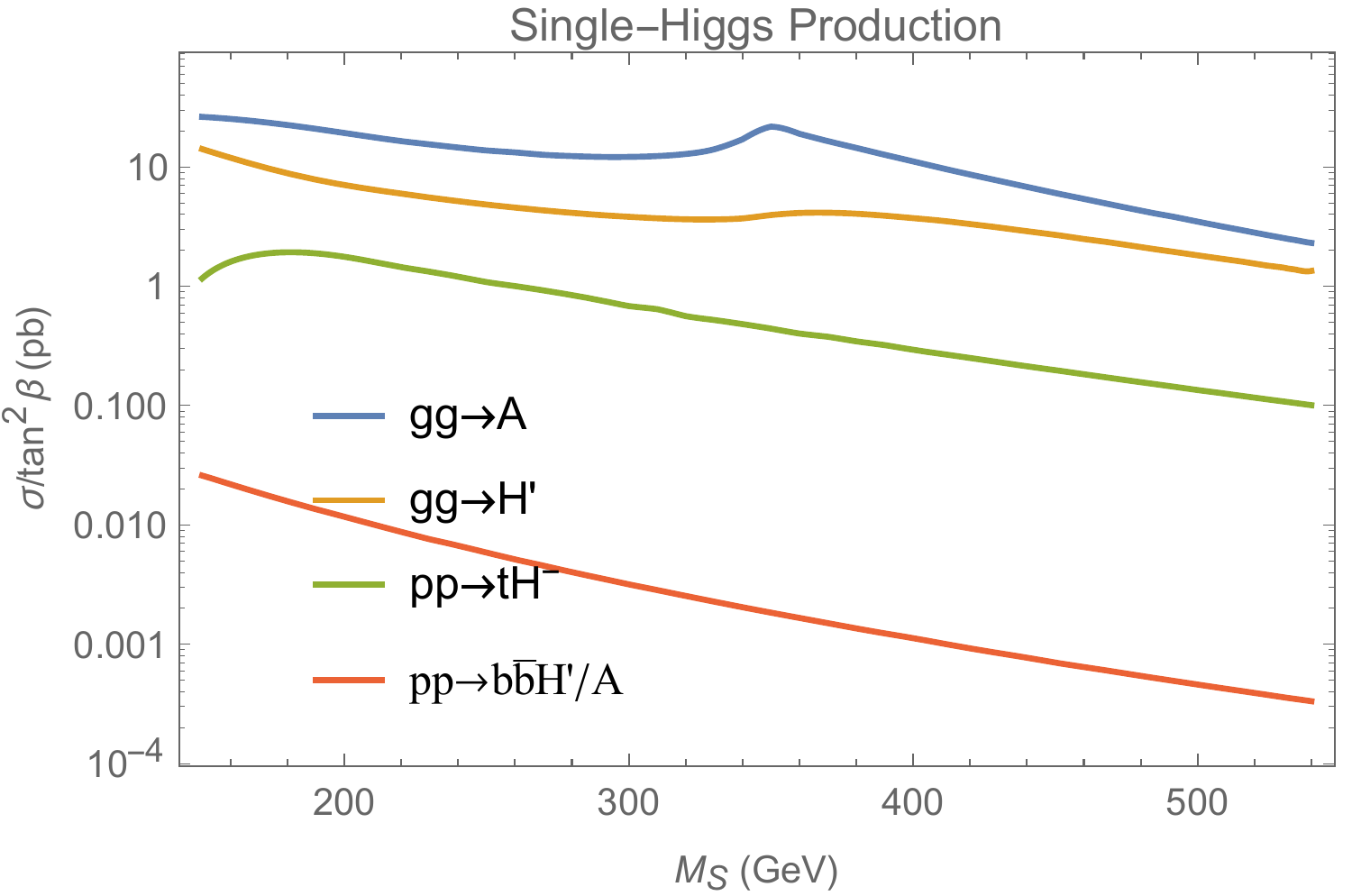}
\caption{The cross sections for $\sqrt{s} = 13\,\tev$ at the LHC for single
  Higgs production processes in the alignment limit ($\delta \to 0$) of the
  GW-2HDM with the dependence on $\tan\beta$ scaled out. Both charged Higgs
  states are included in $pp \to t\bar b H^-$. From Ref.~\cite{Lane:2018ycs}.}
\label{fig:sigmas}
\end{figure}

The principal search modes for the BSM scalars are via gluon fusion:
\bea
\label{eq:Hpmmodes}
gg &\to& H^+ \bar t b \,\,\, {\rm with}\,\,\, H^+ \to t \bar b\,\,\,{\rm
  and} \,\,\, W^+ H';\\
\label{eq:Amodes}
gg &\to& A \to \bar b b,\,\,\,\bar t t \,\,\,{\rm and}\,\,\,ZH';\\
\label{eq:Hpmodes}
gg &\to& H' \to \bar b b,\,\,\bar t t \,\,{\rm and}\,\,ZA,\,W^\pm H^\mp.
\eea
Cross sections for these processes (with the $\tan\beta$ dependence factored
out) are in Fig.~\ref{fig:sigmas}. It is unlikely that an $H'$ or
$A \to \bar b b$ lighter than $2M_t$ can be seen above the QCD
background~\cite{Sirunyan:2018ikr} unless it is accompanied by
$Z \to \ellp\ellm$. If it is, then the $\bar bb\,\ellp\ellm$ signal should be
with reach of ATLAS and CMS capabilities. The decays $H^\pm \to W^\pm H'$ and
$A(H') \to Z H'(A)$ are quickly dominant once the channels open because the
weak boson becomes longitudinally polarized and the decay rate proportional
to $p^3/M_{W,Z}^2$. Thus, they are important near the upper end of the mass
range allowed by the sum rule~(\ref{eq:gensum}).

\begin{itemize}

\item[1)] {\ul{$gg \to H^+\bar tb \to \bar tt \bar bb$}}\hfil\break There
  have been four searches for this process relevant to the mass range of the
  GW-2HDM~\cite{Khachatryan:2015qxa,Aaboud:2018cwk,Sirunyan:2019arl,
    Aad:2021xzu}. The first of these was a CMS search at $8\,\tev$; the other
  three are from 13-TeV data with the last an ATLAS search using the full
  Run~2 data set of $139\,\ifb$. The 8-TeV search by CMS was used to set the
  limit $\tan\beta \simle 0.5$ for $180\,\gev < M_{H^\pm} < 500\,\gev$.
  Subsequent searches have not improved on this limit despite the larger
  date-sets. For example, the limit for the 200--500~GeV mass range extracted
  from Ref.~\cite{Aad:2021xzu} is $\tan\beta < 1.10 \pm 0.14$. The main
  reason for this disappointing outcome is the large $\bar tt$ background at
  low masses and the fact that its rate increases with energy faster than the
  signal does. Given the payoff of a significant improvement in the limit on
  $\tan\beta$, we urge ATLAS and CMS to find a way to overcome the problems
  of searches at low masses.

  There have been no dedicated searches for $gg \to H^+ \bar tb$ with
  $H^+ \to W^+ H' \to W^+ \bar bb$ but this and the main process (without
  resonant $\bar bb$) have the same final state. Hence, $H^+ \to W^+ H'$ may
  unintentionally be included in a search for $H^+ \to t \bar b$.  Even if
  that happened, the model expectation
  $\sigma(gg \to H^+ \bar tb) = 0.075\,\pb$ for $M_{H^\pm} \simeq 400\,\gev$
  and $\tan\beta = 0.5$ is well below the new 95\%~CL ATLAS limit of
  $0.42\,\pb$.

\item[2)] {\ul {$gg \to A/H' \to \bar tt$}}
  \hfil\break A search by CMS with $35.9\,\ifb$ of data at $13\,\tev$ for
  $A/H' \to \bar tt$ with low mass, $400 < M_{A/H'} < 750\,\gev$, is in
  Ref.~\cite{Sirunyan:2019wph}. CMS presented results in terms of allowed and
  excluded regions of the ``coupling
  strength''~$g_{\varphi} = \lambda_{\varphi t\bar t}/(M_t/v)$ and for fixed
  width-to-mass ratio $\Gamma_\varphi/M_\varphi = 0.5$--25\%. In the GW-2HDM,
  $g_{\varphi} = \tan\beta$. For the $C\!P$-odd case, $\varphi = A$, with
  $400\,\gev < M_A < 500\,\gev$ and all $\Gamma_A/M_A$ considered, the region
  $\tan\beta < 0.5$ is not excluded.\footnote{The same appears to be true for
    $\varphi = H'$ with $\Gamma_{H'}/M_{H'} \simge 1\%$.} This is possibly
  due to an excess at $400\,\gev$ that corresponds to a global (local)
  significance of $1.9\,\,(3.5\pm 0.3)\,\sigma$ for
  $\Gamma_A/M_A \simeq 4\%$. The CMS paper notes that higher-order
  electroweak corrections to the SM $t\bar t$ threshold production may
  account for the excess and that further improvement in the theoretical
  description is needed.

\item[3)]
  {\ul{$gg \to A(H') \to Z H'(A) \to \ellp\ellm\,\bar bb$}}\hfil\break There
  have been three published searches so far: \cite{Aaboud:2018eoy,
    Sirunyan:2019wrn,Aad:2020ncx}. The latter ATLAS search updates the former
  one with the full Run~2 data set. As noted above, the decay rates of these
  processes are proportional to $p^3$ and, therefore, they are sensitive to
  the available phase space. They quickly become dominant when
  $M_A = M_{H^\pm} \simge 400\,\gev$ or
  $M_H'\simge 350\,\gev$~\cite{Brooijmans:2020yij, Lane:2019dbc}. Two
  examples of this are shown in Table~\ref{tab:limits} which has been updated
  from Ref.~\cite{Lane:2019dbc} to include the recent ATLAS analysis. The
  full Run~2 data set has significantly improved the ATLAS limits on
  $\sigma B$ which were $255\,\fb$ and $105\,\fb$ for a $36.1\,\ifb$ data set
  in Ref.~\cite{Aaboud:2018eoy}; in particular, the GW-2HDM mass point $M_A =
  M_{H^\pm} = 300\,\gev$, $M_{H'} = 500\,\gev$ is now excluded for $\tan\beta
  = 0.5$; it would be allowed for $\tan\beta < 0.35$.
\begin{table}[!h]
     \begin{center}{
  \begin{tabular}{|c|c|c|c|c|}
  \hline
$M_A = M_{H^\pm}$ & $M_{H'}$ & ATLAS & CMS & GW-2HDM\\
\hline
400 & 300 & 90 &  75 &  65\\
300 & 500 & 51 &  50 & 100\\
\hline
\end{tabular}}
\caption{95\% CL upper limits on
  $\sigma(pp \to A(H'))\, B(A(H') \to ZH'(A)) B(H'(A) \to \bar bb)$ via gluon
  fusion from ATLAS~\cite{Aad:2020ncx}, CMS~\cite{Sirunyan:2019wrn} and
  GW-2HDM calculations for two cases of large $M_A$ or $M_{H'}$. The CMS
  limits include $B(Z \to e^+e^-,\,\mu^+\mu^-)$; the ATLAS limits and GW-2HDM
  predictions do not. Masses are in GeV and $\sigma B$ in~femtobarns.
  $M_A = M_{H^\pm}$ is assumed and $M_{H'}$ is taken from
  Eq.~(\ref{eq:gensum}). Model cross sections are from Fig.~\ref{fig:sigmas}
  with $\tan\beta = 0.50$.}
\label{tab:limits}
\end{center}
\end{table}

\vfil\eject

A search for $A(H') \to Z H'(A) \to \ellp\ellm \bar bb$ using
Contur~\cite{Butterworth:2016sqg} was carried out at the Les Houches~2019
``Physics at TeV colliders'' workshop~\cite{Brooijmans:2020yij}. It showed no
significant sensitivity to these processes for $\tan\beta < 0.5$ except near
$M_A \ge 400\,\gev$ where there was a~$\ge 2\sigma$ exclusion requiring
$\tan\beta \simle 0.3$. This exclusion was based on ATLAS 8-TeV data for
$\ellp\ellm+{\rm jets}$. This study was recently extended using several full
Run~2 ATLAS data sets~(see
{\url{https://hepcedar.gitlab.io/contur-webpage/results/G-W/index.html}}.)
The $1\sigma$ exclusion now covers $\tan\beta < 0.9$ extending down to
$\tan\beta \simeq 0.43$ at $M_A = 150$--$200\,\gev$ and
$\tan\beta \simge 0.23$ at $M_A \simge 400\,\gev$. The greatest sensitivity
came from ATLAS jet and top measurements except near $M_A \ge 400~{\rm GeV}$
where ATLAS 4-lepton measurements dominated.


\end{itemize}

\section*{Acknowledgments}

We are grateful for valuable guidance from Stephen Martin and for our
collaboration with Eric Pilon. We have benefited from discussions on the LHC
data with Tulika Bose, Kevin Black, Gustav Brooijmans, Jon Butterworth,
Viviana Cavaliere, William Murray and David Sperka.


\bibliography{Two-Loop}

\providecommand{\href}[2]{#2}\begingroup\raggedright\begin{thebibliography}{10}

\bibitem{Aad:2012tfa}
{\bf ATLAS} Collaboration, G.~Aad {\em et.~al.}, ``{Observation of a new
  particle in the search for the Standard Model Higgs boson with the ATLAS
  detector at the LHC},'' {\em Phys.Lett.} {\bf B716} (2012) 1--29,
  \href{http://xxx.lanl.gov/abs/1207.7214}{ 1207.7214}.

\bibitem{Chatrchyan:2012ufa}
{\bf CMS} Collaboration, S.~Chatrchyan {\em et.~al.}, ``{Observation of a new
  boson at a mass of 125 GeV with the CMS experiment at the LHC},'' {\em
  Phys.Lett.} {\bf B716} (2012) 30--61,
  \href{http://xxx.lanl.gov/abs/1207.7235}{ 1207.7235}.

\bibitem{Boudjema:2001ii}
F.~Boudjema and A.~Semenov, ``{Measurements of the SUSY Higgs selfcouplings and
  the reconstruction of the Higgs potential},'' {\em Phys. Rev. D} {\bf 66}
  (2002) 095007, \href{http://xxx.lanl.gov/abs/hep-ph/0201219}{
  hep-ph/0201219}.

\bibitem{Gunion:2002zf}
J.~F. Gunion and H.~E. Haber, ``{The CP conserving two Higgs doublet model: The
  Approach to the decoupling limit},'' {\em Phys. Rev.} {\bf D67} (2003)
  075019, \href{http://xxx.lanl.gov/abs/hep-ph/0207010}{ hep-ph/0207010}.

\bibitem{Carena:2013ooa}
M.~Carena, I.~Low, N.~R. Shah, and C.~E.~M. Wagner, ``{Impersonating the
  Standard Model Higgs Boson: Alignment without Decoupling},'' {\em JHEP} {\bf
  04} (2014) 015, \href{http://xxx.lanl.gov/abs/1310.2248}{ 1310.2248}.

\bibitem{Haber:2018ltt}
H.~E. Haber, ``{Approximate Higgs alignment without decoupling},'' in {\em
  {53rd Rencontres de Moriond on QCD and High Energy Interactions}},
  pp.~139--142.
\newblock 2018.
\newblock \href{http://xxx.lanl.gov/abs/1805.05754}{ 1805.05754}.

\bibitem{Ivanov:2007de}
I.~P. Ivanov, ``{Minkowski space structure of the Higgs potential in 2HDM. II.
  Minima, symmetries, and topology},'' {\em Phys. Rev. D} {\bf 77} (2008)
  015017, \href{http://xxx.lanl.gov/abs/0710.3490}{ 0710.3490}.

\bibitem{Dev:2014yca}
P.~S. Bhupal~Dev and A.~Pilaftsis, ``{Maximally Symmetric Two Higgs Doublet
  Model with Natural Standard Model Alignment},'' {\em JHEP} {\bf 12} (2014)
  024, \href{http://xxx.lanl.gov/abs/1408.3405}{ 1408.3405}. [Erratum:
  JHEP11,147(2015)].

\bibitem{Benakli:2018vqz}
K.~Benakli, M.~D. Goodsell, and S.~L. Williamson, ``{Higgs alignment from
  extended supersymmetry},'' {\em Eur. Phys. J.} {\bf C78} (2018), no.~8, 658,
  \href{http://xxx.lanl.gov/abs/1801.08849}{ 1801.08849}.

\bibitem{Darvishi:2019ltl}
N.~Darvishi and A.~Pilaftsis, ``{Quartic Coupling Unification in the Maximally
  Symmetric 2HDM},'' {\em Phys. Rev. D} {\bf 99} (2019), no.~11, 115014,
  \href{http://xxx.lanl.gov/abs/1904.06723}{ 1904.06723}.

\bibitem{Gildener:1976ih}
E.~Gildener and S.~Weinberg, ``{Symmetry Breaking and Scalar Bosons},'' {\em
  Phys. Rev.} {\bf D13} (1976) 3333.

\bibitem{Coleman:1973jx}
S.~R. Coleman and E.~J. Weinberg, ``{Radiative Corrections as the Origin of
  Spontaneous Symmetry Breaking},'' {\em Phys. Rev.} {\bf D7} (1973)
  1888--1910.

\bibitem{Lee:2012jn}
J.~S. Lee and A.~Pilaftsis, ``{Radiative Corrections to Scalar Masses and
  Mixing in a Scale Invariant Two Higgs Doublet Model},'' {\em Phys. Rev.} {\bf
  D86} (2012) 035004, \href{http://xxx.lanl.gov/abs/1201.4891}{ 1201.4891}.

\bibitem{Glashow:1976nt}
S.~L. Glashow and S.~Weinberg, ``{Natural Conservation Laws for Neutral
  Currents},'' {\em Phys. Rev.} {\bf D15} (1977) 1958.

\bibitem{Lane:2018ycs}
K.~Lane and W.~Shepherd, ``{Natural stabilization of the Higgs boson’s mass
  and alignment},'' {\em Phys. Rev.} {\bf D99} (2019), no.~5, 055015,
  \href{http://xxx.lanl.gov/abs/1808.07927}{ 1808.07927}.

\bibitem{Branco:2011iw}
G.~C. Branco, P.~M. Ferreira, L.~Lavoura, M.~N. Rebelo, M.~Sher, and J.~P.
  Silva, ``{Theory and phenomenology of two-Higgs-doublet models},'' {\em Phys.
  Rept.} {\bf 516} (2012) 1--102, \href{http://xxx.lanl.gov/abs/1106.0034}{
  1106.0034}.

\bibitem{Khachatryan:2015qxa}
{\bf CMS} Collaboration, V.~Khachatryan {\em et.~al.}, ``{Search for a charged
  Higgs boson in pp collisions at $ \sqrt{s}=8 $ TeV},'' {\em JHEP} {\bf 11}
  (2015) 018, \href{http://xxx.lanl.gov/abs/1508.07774}{ 1508.07774}.

\bibitem{Aaboud:2018cwk}
{\bf ATLAS} Collaboration, M.~Aaboud {\em et.~al.}, ``{Search for charged Higgs
  bosons decaying into top and bottom quarks at $\sqrt{s}$ = 13 TeV with the
  ATLAS detector},'' {\em JHEP} {\bf 11} (2018) 085,
  \href{http://xxx.lanl.gov/abs/1808.03599}{ 1808.03599}.

\bibitem{Lane:2019dbc}
K.~Lane and E.~Pilon, ``{Phenomenology of the new light Higgs bosons in
  Gildener-Weinberg models},'' {\em Phys. Rev. D} {\bf 101} (2020), no.~5,
  055032, \href{http://xxx.lanl.gov/abs/1909.02111}{ 1909.02111}.

\bibitem{Martin:2001vx}
S.~P. Martin, ``{Two Loop Effective Potential for a General Renormalizable
  Theory and Softly Broken Supersymmetry},'' {\em Phys. Rev.} {\bf D65} (2002)
  116003, \href{http://xxx.lanl.gov/abs/hep-ph/0111209}{ hep-ph/0111209}.

\bibitem{Jackiw:1974cv}
R.~Jackiw, ``{Functional evaluation of the effective potential},'' {\em Phys.
  Rev. D} {\bf 9} (1974) 1686.

\bibitem{Hashino:2015nxa}
K.~Hashino, S.~Kanemura, and Y.~Orikasa, ``{Discriminative phenomenological
  features of scale invariant models for electroweak symmetry breaking},'' {\em
  Phys. Lett.} {\bf B752} (2016) 217--220,
  \href{http://xxx.lanl.gov/abs/1508.03245}{ 1508.03245}.

\bibitem{Braathen:2020vwo}
J.~Braathen, S.~Kanemura, and M.~Shimoda, ``{Two-loop analysis of classically
  scale-invariant models with extended Higgs sectors},'' {\em JHEP} {\bf 03}
  (2021) 297, \href{http://xxx.lanl.gov/abs/2011.07580}{ 2011.07580}.

\bibitem{Battye:2011jj}
R.~A. Battye, G.~D. Brawn, and A.~Pilaftsis, ``{Vacuum Topology of the Two
  Higgs Doublet Model},'' {\em JHEP} {\bf 08} (2011) 020,
  \href{http://xxx.lanl.gov/abs/1106.3482}{ 1106.3482}.

\bibitem{Pilaftsis:2011ed}
A.~Pilaftsis, ``{On the Classification of Accidental Symmetries of the Two
  Higgs Doublet Model Potential},'' {\em Phys. Lett.} {\bf B706} (2012)
  465--469, \href{http://xxx.lanl.gov/abs/1109.3787}{ 1109.3787}.

\bibitem{Sirunyan:2018ikr}
{\bf CMS} Collaboration, A.~M. Sirunyan {\em et.~al.}, ``{Search for low-mass
  resonances decaying into bottom quark-antiquark pairs in proton-proton
  collisions at $\sqrt{s} =$ 13 TeV},'' {\em Phys. Rev.} {\bf D99} (2019),
  no.~1, 012005, \href{http://xxx.lanl.gov/abs/1810.11822}{ 1810.11822}.

\bibitem{Sirunyan:2019arl}
{\bf CMS} Collaboration, A.~M. Sirunyan {\em et.~al.}, ``{Search for a charged
  Higgs boson decaying into top and bottom quarks in events with electrons or
  muons in proton-proton collisions at $ \sqrt{\mathrm{s}} $ = 13 TeV},'' {\em
  JHEP} {\bf 01} (2020) 096, \href{http://xxx.lanl.gov/abs/1908.09206}{
  1908.09206}.

\bibitem{Aad:2021xzu}
{\bf ATLAS} Collaboration, G.~Aad {\em et.~al.}, ``{Search for charged Higgs
  bosons decaying into a top quark and a bottom quark at $\sqrt{s}$=13 TeV with
  the ATLAS detector},'' \href{http://xxx.lanl.gov/abs/2102.10076}{
  2102.10076}.

\bibitem{Sirunyan:2019wph}
{\bf CMS} Collaboration, A.~M. Sirunyan {\em et.~al.}, ``{Search for heavy
  Higgs bosons decaying to a top quark pair in proton-proton collisions at
  $\sqrt{s} =$ 13 TeV},'' {\em JHEP} {\bf 04} (2020) 171,
  \href{http://xxx.lanl.gov/abs/1908.01115}{ 1908.01115}.

\bibitem{Aaboud:2018eoy}
{\bf ATLAS} Collaboration, M.~Aaboud {\em et.~al.}, ``{Search for a heavy Higgs
  boson decaying into a $Z$ boson and another heavy Higgs boson in the
  $\ell\ell bb$ final state in $pp$ collisions at $\sqrt{s}=13$ TeV with the
  ATLAS detector},'' {\em Phys. Lett.} {\bf B783} (2018) 392--414,
  \href{http://xxx.lanl.gov/abs/1804.01126}{ 1804.01126}.

\bibitem{Sirunyan:2019wrn}
{\bf CMS} Collaboration, A.~M. Sirunyan {\em et.~al.}, ``{Search for new
  neutral Higgs bosons through the H$\to$ ZA $\to \ell^{+}\ell^{-}
  \mathrm{b\bar{b}}$ process in pp collisions at $\sqrt{s} =$ 13 TeV},'' {\em
  JHEP} {\bf 03} (2020) 055, \href{http://xxx.lanl.gov/abs/1911.03781}{
  1911.03781}.

\bibitem{Aad:2020ncx}
{\bf ATLAS} Collaboration, G.~Aad {\em et.~al.}, ``{Search for a heavy Higgs
  boson decaying into a Z boson and another heavy Higgs boson in the $\ell \ell
  bb$ and $\ell \ell WW$ final states in $pp$ collisions at
  $\sqrt{s}=13$~$\text {TeV}$ with the ATLAS detector},'' {\em Eur. Phys. J. C}
  {\bf 81} (2021), no.~5, 396, \href{http://xxx.lanl.gov/abs/2011.05639}{
  2011.05639}.

\bibitem{Brooijmans:2020yij}
G.~Brooijmans {\em et.~al.}, ``{Les Houches 2019 Physics at TeV Colliders: New
  Physics Working Group Report},''
\newblock 2020.
\newblock \href{http://xxx.lanl.gov/abs/2002.12220}{ 2002.12220}.

\bibitem{Butterworth:2016sqg}
J.~M. Butterworth, D.~Grellscheid, M.~Kr\"amer, B.~Sarrazin, and D.~Yallup,
  ``{Constraining new physics with collider measurements of Standard Model
  signatures},'' {\em JHEP} {\bf 03} (2017) 078,
  \href{http://xxx.lanl.gov/abs/1606.05296}{ 1606.05296}.

\end{thebibliography}\endgroup
\bibliographystyle{utcaps}
\end{document}